# Penetration through a wall: Is it reality?

B. Ivlev

**A tennis ball is not expected to penetrate through a brick wall since a motion under a barrier is impossible in classical mechanics. With quantum effects a motion of a particle through a barrier is allowed due to quantum tunneling. According to usual theories of tunneling, the particle density decays inside a classical barrier resulting in an extremely slow penetration process [1–4]. However, there are no general laws forbidding fast motion through classical barriers. The problem addressed is investigation of unusual features of quantum tunneling through a classical static barrier which is at least two-dimensional. Here we show that penetration through such barrier can be not slow [5]. When the barrier satisfies the certain conditions, a regime of quantum lens is possible with formation of caustics. De Broglie waves are reflected from the caustics, interfere, and result in a not small flux from under the barrier. This strongly contrasts to the usual scenario with a decaying under-barrier density [1–4]. We construct a particular example of fast motion through a classical barrier. One can unexpectedly conclude that, in principle, nature allows fast penetration through classical barriers which is against common sense. The phenomenon may be responsible for a variety of processes in labs and nature. For example, tunneling in solids may occur with a different scenario, in biophysics and chemistry one can specify conditions for unusual reactions, and evanescent optical waves may strongly change their properties. In condensed matter and cosmic physics there are phenomena with mysterious reasons of an energy emission, for instance, gamma-ray bursts [6]. One can try to treat them in the context of fast escape from under some barriers.**

In everyday life a wall is an absolute obstacle. In the world of electrons, walls are not so hard due to quantum tunneling [7–10]. There are criteria of weak penetration through a barrier, for example, if it is classical (high and wide) [4]. In this case tunneling across the barrier is extremely slow. However, under certain conditions fast tunneling can occur when only a slow leakage through the barrier is expected according to the usual scenario. The phenomenon is outside the current opinion formulated in 20$^{\text{th}}$ of the last century [1–3].

The object we study is a straight and long quantum wire (a diameter of a few nanometers) in a static electric field perpendicular to it. The wire and the potential energy of an electron are shown in Fig. 1a. The electron can move free along the wire inside the potential well. This motion is indicated by the red arrows in Fig. 1a. But the perpendicular motion is restricted by the walls resulting in discrete energy levels. The certain impurity is placed outside the wire. The electron energy is a position of the discrete level in the wire plus $mv_Y^2/2$.

For the beginning, suppose that the impurity is absent. In this case motions in the $X$ and $Y$ directions are independent. The motion in the $X$ direction is forbidden by the Newtonian mechanics. In quantum physics it is allowed as tunneling and occurs through the tilted barrier in Fig. 1a formed by the applied electric field. The decay of the electron density under the barrier is depicted in Fig. 1b. It is generic with a usual decay in a one-dimensional barrier described in textbooks[4]. Directions of electron velocity are shown in Fig. 1c. The $Y$ component of the velocity does not depend on coordinates but the $X$ component increases with $X$ and can become comparable with the $Y$ component. In contrast to the velocity, the current under the barrier is small since it is a product of the velocity and a particle density.

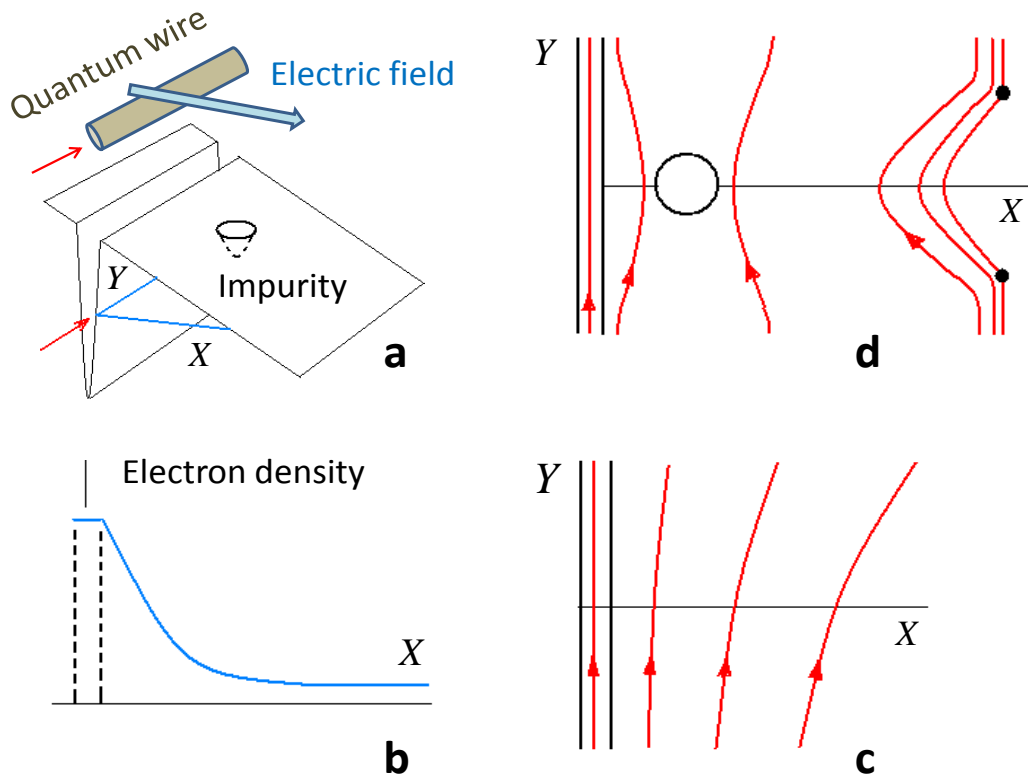

**Figure 1 | Features of the barrier penetration. a,** Quantum wire, placed in the $Y$ direction, and the electron potential energy in the two-dimensional space. The constant electric field $\mathcal{E}$ is directed perpendicular to the wire. The impurity is placed at the barrier region. The electron velocity $v_Y$ along the wire is shown by the red arrows. **b,** The spatial decay of the electron density under the barrier in the absence of the impurity. The region between dashed lines is the interior of the wire. **c,** Velocity directions of the electron in the absence of the impurity. **d,** Velocity directions of the electron. The impurity is depicted by the circle. The singularities in the velocity distribution are shown by the black dots.

Let us return the impurity to its position in Fig. 1a. Along the impurity, extended in space, the phase of the de Broglie wave of the electron is strongly accumulated. The large phase allows to consider the electron state as one consisted of classical rays. The analogous approach for Maxwell equations is called geometrical optics[11]. Rays of de Broglie waves are Newtonian trajectories. The distribution of their velocities is illustrated in Fig. 1d. Close to the impurity the electron is attracted to it since it has a lower energy. But at a larger distance an unexpected phenomenon occurs. The velocity exhibits singularities marked in Fig. 1d by the black points. In terms of rays in geometrical optics, at the singularity point the velocity vector jumps in its direction. In a close vicinity of the singularity the geometrical optics is not applicable going over into wave optics when one cannot neglect a short wave length. As a result, on the short scale the exact electron velocity is not singular. Note that the singularities arise only when the velocity along the wire is finite. Otherwise it would be a usual barrier penetration with a decaying density of the type shown in Fig. 1b.

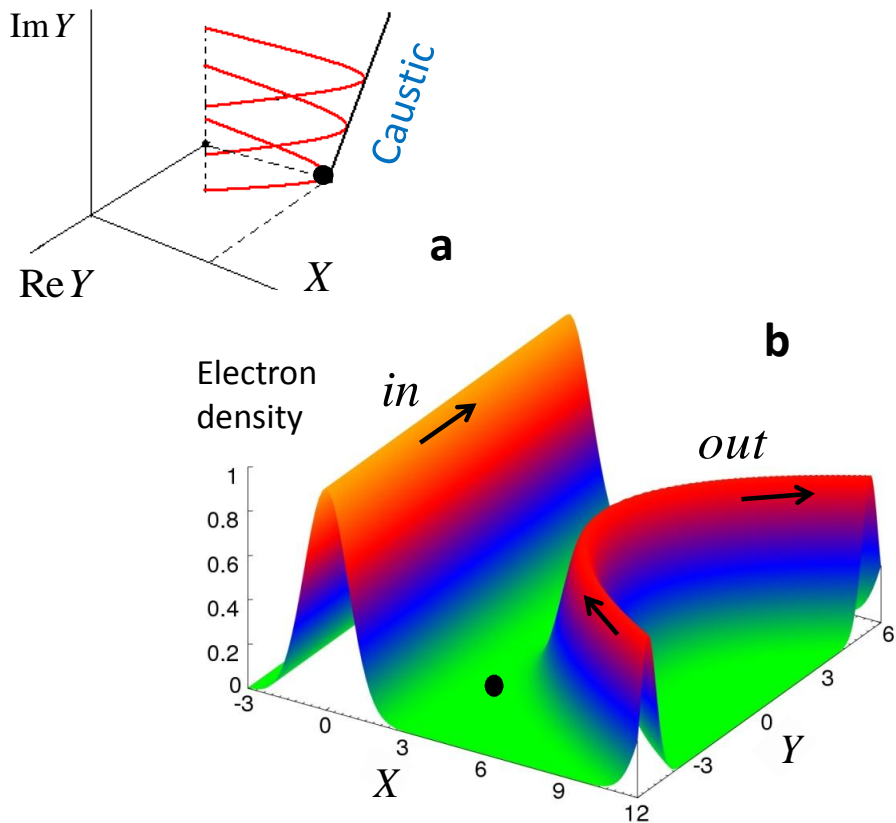

**Figure 2 | Singularities in the electron system. a,** Reflection of complex Newtonian trajectories from the caustic. The caustic pierces the physical plane at the point marked by the black dot. **b,** Spatial dependence of the electron density in arbitrary units. The part $in$ is localized at the quantum wire. The black dot marks the singular point where the caustic pierces the physical plane. The other dot is invisible. The part $out$ is maximum on the classical trajectory. The arrows indicate velocity directions.

To clarify the above features one should go to a more detailed description in terms of an electron wave function $\psi(X,Y)$ which obeys equations of quantum mechanics [4]. At a large $X$ in Fig. 1a the classical motion of the electron occurs along the trajectory

$$X = X_0 + \frac{\varepsilon}{2mv_Y^2}(Y-Y_0)^2, \qquad (1)$$

which relates to a reflection from the barrier and follows from the Newtonian mechanics. In terms of rays (trajectories) of geometrical optics it is convenient to consider the wave function not in the entire two-dimensional plane but on trajectories of the type (1), $\psi[X(Y),Y]$. To penetrate into the classically forbidden region under the barrier, $X < X_0$, one can formally put imaginary $Y - Y_0$ in equation (1). As a result, the classical trajectory becomes complex. This is a famous method to consider a motion which is impossible in the Newtonian mechanics [12–15]. To describe such trajectories one has to extend dimensionality of the space up to three as shown in Fig. 2a. It happens that the wave function $\psi(X,Y)$, where $Y$ is now a complex variable, varies rapidly at a close vicinity of the certain curve in the tree-dimensional space. At that region the length of the de Broglie wave is not short compared to the size of the region. This means that geometrical optics breaks down close to that singular curve which is "thick". The curve pierces the physical plane $\{X, \mathrm{Re}\,Y\}$ at the point of a singularity depicted in Fig. 1d. Each trajectory (1) relates to the certain constants $X_0$ and $Y_0$ and is reflected from that curve called caustic [11] as shown in Fig. 2a. So the system behaves as quantum lens. Caustics came to quantum mechanics [16–19,5] from optics.

In our problem the impurity is extended. For simplicity we place it directly at the wire [5] so that the position of the discrete level, counted from the barrier top, was proportional to $[1 + 0.03\exp(-Y^2/a^2)]^{1/2}$. This differs from the point impurity in paper [20]. The spatial scale $a$ was specified by the condition when $a\varepsilon$ is of the order of the discrete energy level in the wire [5]. The condition was chosen that $mv_Y^2/2$ is 0.2 of the level position away from the impurity. The resulting electron density is shown in Fig. 2b. The part *in* looks as a usual electron density localized at the wire. The black dot is the singularity where the caustic pierces the plane. The top of the part *out* corresponds to the classical motion (1) with $Y_0 = 0$. There is no a net particle flux in the $X$ direction. The $Y$ component of the velocity is finite in the both parts, *in* and *out*.

There are two independent states in the problem. The first corresponds to an outer particle which collides the barrier and reflects. The second is associated with the wire.

Without the impurity it is similar to the part *in* in Fig. 2b. With the impurity the wire-associated state is quite different since it consists of the parts *in* and *out* which are not independent. The part *out* is generated due to the singularities [5] and is a superposition (with a fixed total energy) of $Y$ waves and incident-reflected $X$ waves with different velocities.

In motion around the singular point in Fig. 2b an electron state, corresponding to geometrical optics, goes over into a different state also related to geometrical optics which can describe each state separately but not local conversion of them [21,22]. Therefore geometrical optics can be not sufficient to study a barrier penetration by a particle with a velocity transverse to the tunneling direction. The analysis of conversion of states in such problems is delicate [21,22]. Nevertheless, there is the certain feature of our phenomenon which allows to easily conclude about the outer branch. Namely, one can move, from the position *in* to the position *out* in Fig. 2b, between the two singularities with no violation of geometrical optics and being within the same branch. Therefore the electron density in Fig. 2b, obtained by the methods of geometrical optics, is a correct representation of the exact solution of quantum mechanical equations [5].

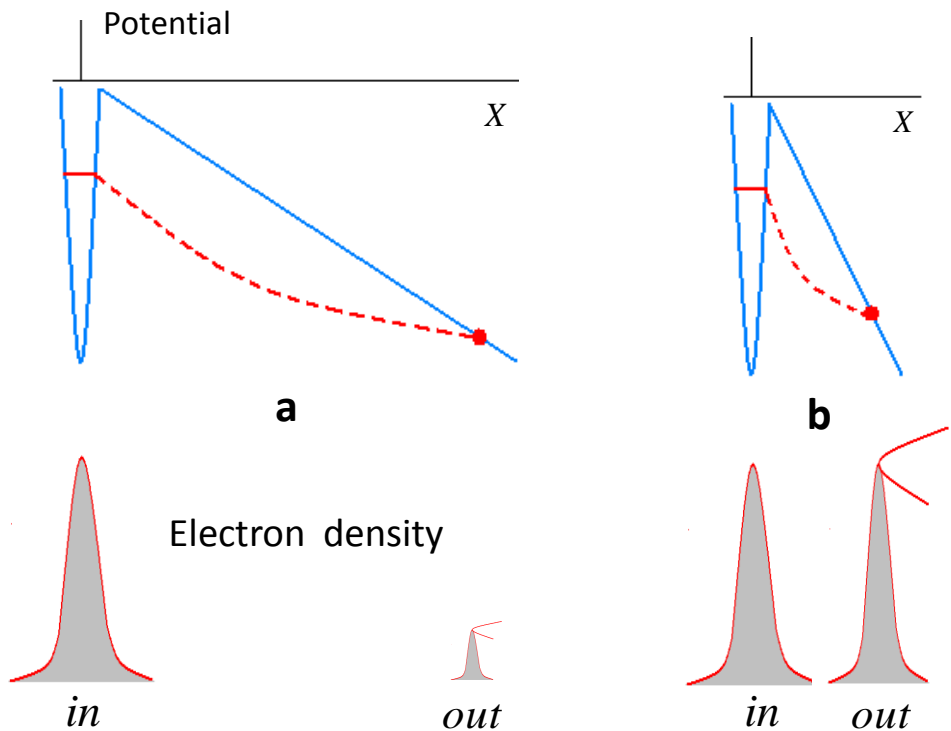

**Figure 3 | Barrier penetration for different electric fields. a,** Potential barrier of the wire and the electron density at $Y=0$ for a weak electric field. **b,** The same at a large electric field. The densities *in* and *out* are equal in amplitude.

The part *out* increases with the electric field, as shown in Fig. 3, and becomes equal in amplitude to the part *in* when $a\varepsilon$ is 1.60 of the discrete level position in the wire[5]. This is shown in Fig. 3b. Note, that at this electric field the barrier is still not transparent in the conventional description.

Suppose that at the moment $t = 0$ the electron density is only the part *in* in Fig. 2b and the part *out* is removed as shown in Fig. 4a. According to usual theories, this state is almost static (metastable) since only a weak time dependence is expected due to small under-barrier tails. The state is a superposition of the full one in Fig. 2(b) and over-barrier waves which compensate the part *out*. The over-barrier states enter with weights $\exp(-iEt/\hbar)$ and do not compensate the part *out* at a finite $t$. Therefore outside the barrier (empty region in Fig. 4a) at a positive $t$ the electron density becomes finite. The dynamics is fast with the time involved inversely proportional to the barrier height.

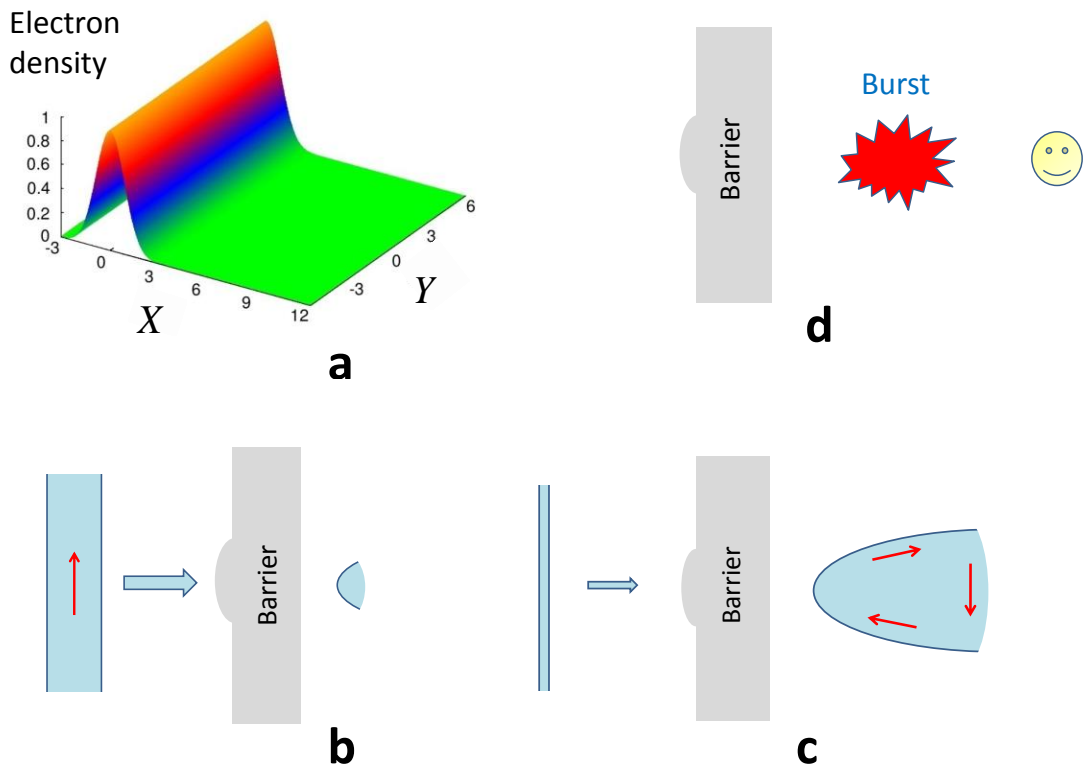

**Figure 4 | Dynamics of the barrier penetration. a,** The initial electron state is generic with the part *in* in Fig. 2b. **b,** Beginning of the decay of the pre-barrier state whose velocity is marked by the arrow. The part *out* starts to be formed. The non-homogeneous barrier is shown schematically. **c,** The same for a later moment of time. The velocities in the expanding domain to the right of the barrier are depicted by the arrows. **d,** Barrier penetration with dissipation to the right of it. The outgoing part emits energy received by the observer shown to the right.

In other words, the state in the wire (Fig. 4a) is blown up when the electric field reaches the value related to Fig. 3b. The empty region starts to get quickly filled out despite the state in the wire (Fig. 4a) is treated as long living one by conventional theories. *This constitutes the process of fast penetration through classical barriers.* At lower fields (Fig. 3a) the state in Fig. 4a lives a long time as a metastable one.

Fast penetration can start only from a state with a finite transverse kinetic energy $mv_Y^2/2$. At zero transverse velocity we deal with a usual long living state. From the usual standpoint, all states in the wire (Fig. 1a) with energies below the barrier are long living ones. In contrast, according to the above mechanism, states in the well with finite velocities along the wire can be destroyed by the fast transfer outside the barrier. Those states live long only at a small electric field. Therefore a state, considered as long living, can suddenly decay fast if to slow increase the electric field. We emphasize that the barrier remains non-transparent if to adopt a conventional scenario of tunneling.

The fast dynamics of formation of the outer part from the initial pre-barrier density is shown in Figs. 4b and 4c. The system tends to recover the wave function in Fig. 2b. We consider a non-dissipative problem when the outgoing flux carries the energy stored at the pre-barrier region. With dissipation, the outgoing flux emits energy to a surround. From the standpoint of the observer to the right of the barrier, depicted in Fig. 4d, this process looks as a formation of some burst which emits energy from "nothing" since the observer does not expect a strong energy flux from the barrier. Burst formation can occur under variation of barrier parameters (increasing electric field in the case of the wire). If there is a permanent particle flux to the barrier from the left in Fig. 4, the observer in Fig. 4d will see a permanent energy flux from "nothing" which reminds some energy generator.

The method of study is based on a solution of quantum mechanical equations in two dimensions described in details in paper [5]. The approach is analogous to the limit of geometrical optics for Maxwell equations. In our problem this solution can be obtained exactly. The geometrical optics is violated in vicinities of local singularities where the caustics pierce the physical plane. Within the geometrical optics, the singularities serve as branching points and one should use wave optics in their vicinities to select correct branches. A direct numerical analysis of quantum mechanical equations requires an extremely high precision to resolve small branches and is impossible now [5].

We consider the problem when classical physics is not applicable to a phenomenon which is obviously expected to be practically classic. The electron density initially decays inside the classical barrier (the part *in* in Fig. 2b) as in a usual tunneling problem. At first sight, this should result in an extremely small output from under the barrier which is zero in classical physics. But formation of the part *out* at large distances completely destroys this scenario and leads to a strong penetration through the classical barrier.

It is hard to list all manifestations of the phenomenon since a barrier penetration is a part of a variety of natural and artificial processes[7–10,23–27]. Here we consider an example in condensed matter physics with an escape from a non-homogeneous quantum wire. Artificial junctions of metals or semiconductors, separated by classical barriers and being properly adjusted, can exhibit the unusual penetration of electrons. In field emission experiments (tunneling of electrons to vacuum from a material in an applied electric field) one can also observe fast penetration phenomenon under certain conditions, for example, a non-flat material surface. Tunneling of an elastic string in a tilted wash-board potential[28] models various phenomena. See also[29]. The process can be strongly enhanced by a longitudinal velocity of the string and an inhomogeneity along the board. It is amazing that an exotic phenomenon of ball lightning[30] reminds the scenario in Fig. 4d. A mechanism of the energy pump to the ball is still unclear.

In conclusion, an unusual phenomenon of penetration through classical barriers is reported. A particle energy is well below the barrier and the Newtonian mechanics forbids such processes. In contrast to the usual quantum mechanism, which allows only an extremely slow leakage through classical barriers, the reported phenomenon is fast penetration through them. Despite the problem studied relates to the microscopic (electrons) level it reveals a principal possibility in nature of penetration through classical barriers of various scales. So the question in the title may not sound strange.

**Acknowledgements** I thank G. Blatter, M. I. Dykman, V. B. Geshkenbein, S. A. Gurvitz, R. S. Thompson, and A. V. Ustinov for discussions. I am also grateful to participants of the seminars at ETH, Zürich, Switzerland and University of Karlsruhe, Karlsruhe, Germany. The research was supported by grant U48369–F CONACYT.

1. Wentzel, G. Eine Verallgemeinerung der Quantenbedingungen für die Zwecke der Wellenmechanik. *Z. Physik* **38**, 518-529 (1926).
2. Kramers, H. A. Wellenmechanik und halbzälige Quantisierung. *Z. Physik* **39**, 828-840 (1926).
3. Brillouin, L. La mécanique ondulatoire de Schrödinger: une méthode générale de resolution par approximations successives. *Comptes Rendus de l'Academie des Sciences* **183**, 24-26 (1926).
4. Landau, L. D. & Lifshitz, E. M. *Quantum Mechanics* (Pergamon, 1977).
5. Ivlev, B. Underbarrier interference. *Ann. Phys*. **326**, 979-1001 (2011).
6. Lyutikov, M. Magnetism in a cosmic blast. *Nature* **462**, 728-729 (2009).

7.  Leggett, A. J. in *Tunneling in Complex Systems* (ed Tosovic, S.) 1-34 (World Scintific, 1998).
8.  Heller, E. J. The Many Faces of Tunneling. *J. Phys. Chem. A* **103**, 10433-10444 (1999).
9.  Ankerhold, J. *Quantum Tunneling in Complex Systems* (Springer-Verlag, 2007).
10. Miller, W. H. Semiclassical limit of quantum mechanical transition state theory for nonseparable systems. *J. Chem. Phys*. **62**, 1899-1906 (1975).
11. Landau, L. D. & Lifshitz, E. M. *The Classical Theory of Fields* (Butterworth-Heinemann, 1998).
12. Stone, M. Semiclassical methods for unstable states. *Phys. Lett*. **67B**, 186-188 (1977).
13. Callan, C. G. & Coleman, S. Fate of the false vacuum. II. First quantum corrections. *Phys. Rev. D* **16**, 1762-1768 (1977).
14. Gervais, J. –L. & Sakita, B. WKB wave function for systems with many degrees of freedom: A unified view of solitons and pseudoparticles. *Phys. Rev. D* **16**, 3507-3514 (1977).
15. Eckern, U. & Schmid, A. in *Quantum Tunneling in Condensed Media* (eds Kagan, Yu. & Leggett, A. J.) 145-229 (Elsevier, 1992).
16. Keller, J. B. Corrected bohr-sommerfeld quantum conditions for nonseparable systems. *Ann. Phys*. **4**, 180-188 (1958).
17. Baz, A. I., Zeldovich, Ya. B. & Perelomov, A. M. *Scattering Reactions and Decay in Nonrelativistic Quantum Mechanics* (Israel Program for Scientific Translations, 1969).
18. Maitra, N. T. & Heller, E. J. Semiclassical amplitudes: Supercaustics and the whisker map. *Phys. Rev. A* **61**, 012107 (1999).
19. Sharpee, T., Dykman, M. I. & Platzman, P. M. Tunneling decay in a magnetic field. *Phys. Rev. A* **65**, 032122 (2002).
20. Vaishnav, J. Y., Itsara, A. & Heller, E. J. Hall of mirrors scattering from an impurity in a quantum wire. *Phys. Rev. B* **73**, 115331 (2006).
21. Heading, J. *An Introduction to Phase-Integral Methods* (John Wiley, 1962).
22. Maslov, V. P. & Fedoriuk, M. V. *Semi-Classical Approximation in Quantum Mechanics* (Kluwer, Academic, 2002).
23. Ivlev, B. & Palomarez-Báez, J. P. Two-dimensional tunneling in a SQUID. *Phys. Rev. B* **82**, 184513 (2010).
24. Weiss, U. *Series in Modern Condensed Matter Physics*, vol. 2 (World Scientific, 1993).
25. Gurvitz, S. A. & Levinson, Y. B. Resonant reflection and transmission in a conducting channel with a single impurity. *Phys. Rev. B* **47**, 10578-10587 (1993).

26. Thorwart, M., Grifoni, M. & Hänggi, P. Strong coupling theory for driven tunneling and vibrational relaxation. *Phys. Rev. Lett*. **85**, 860-863 (2000).
27. Thomann, A. U., Geshkenbein, V. B. & Blatter, G. Quantum instability in a dc SQUID with strongly asymmetric dynamical parameters. *Phys. Rev. B* **79**, 184515 (2009).
28. Ivlev, B. I. & Mel'nikov, V. I. Tunneling and activated motion of a string across a potential barrier. *Phys. Rev. B* **36**, 6889-6903 (1987).
29. Hong, J., Vilenkin, A. & Winitzki, S. Particle creation in a tunneling universe. I. *Phys. Rev. D* **68**, 023520 (2003).
30. Meshcheryakov, O. Ball Lightning-Aerosol Electrochemical Power Source or A Cloud of Batteries. *Nanoscale Res. Lett.* **2**, 319-330 (2007).